# Quantification of phase instabilities in amplitude swing for ultrashort pulse train characterization

**Cristian Barbero,**[1,2,*] **Íñigo J. Sola,**[1,2] **and Benjamín Alonso**[1,2]

[1]*Grupo de Investigación en Aplicaciones del Láser y Fotónica (ALF), Universidad de Salamanca, 37008 Salamanca, Spain*
[2]*Unidad de Excelencia en Luz y Materia Estructuradas (LUMES), Universidad de Salamanca, 37008 Salamanca, Spain*
**cristianbp@usal.es*

**Abstract:** The temporal dynamics of ultrashort pulses are a fundamental feature in ultrafast optics. These dynamics can often be extracted from a two-dimensional trace consisting of a set of nonlinear spectra, using an iterative algorithm. Typically, the measurement of this trace requires integrating the signal of many pulses, which implies that the trace does not correspond to a single pulse when shot-to-shot variations occur. In this case, the pulse train can be characterized by a base pulse and a metric that quantifies its instabilities. Here, we demonstrate that the amplitude swing technique is highly sensitive to pulse-train instabilities, allowing not only for their detection but, crucially, for their quantification. First, we examine the analytical components of the amplitude swing trace. We then introduce a robust parameter designed to assess the instabilities, evaluating its performance in the presence of noise. Finally, we validate this approach using simulated unstable pulse trains, introducing random spectral phase fluctuations, establishing amplitude swing as a simple yet powerful diagnostic tool for full pulse-train characterization.

## 1. Introduction

The measurement of the temporal amplitude and phase of an ultrashort laser pulse is a fundamental, non-trivial task when dealing with such short events. For this purpose, there is a group of techniques based on encoding the temporal dynamics of the pulse into a two-dimensional map, called trace. This trace often represents the spectrum of a nonlinear signal generated by the tested pulse as a function of a technique-specific parameter, i.e., is the nonlinear intensity as a function of the optical frequency and this parameter. Some representative examples of these techniques are FROG [1], d-scan [2], FROSt [3], and a-swing [4], in which the varying parameter are the delay between both replicas, the dispersion introduced by a previous system, the delay between the pulse and an ultrafast switch, and the relative amplitude between two delayed replicas, respectively. Under ideal conditions, each pulse yields a unique trace, and each trace corresponds to a unique pulse. Thus, from the 2D trace, the pulse amplitude and phase are retrieved using an iterative algorithm, that can be based on projections [5,6], gradient-descent optimization [7–9], evolutionary strategies [10,11], ptychography [12–15], neural networks [16,17], or others [18].

The experimental implementation of these techniques usually involves mechanical scanning to acquire the nonlinear signal while varying the corresponding parameter. Therefore, the detector integrates the signals from many pulses, yielding a trace that is the incoherent superposition of the individual traces of the pulses within the train. Thus, the measured trace corresponds to a single pulse only if all pulses exhibit the same amplitude and phase, which is generally a reasonable assumption in pulse characterization.

However, numerous situations exist in which the temporal profile varies from shot to shot, such as supercontinuum lasers [19]. Under these conditions, the measured trace results from contributions of multiple pulses rather than a single pulse, and thus the retrieval admits no unique solution. There are two approaches to address this: the individual measurement of all pulses within the train or the extraction of statistical information regarding the pulse train. The first approach implies the use of single-shot implementations of these characterization techniques, in which the varying parameter is encoded in a spatial dimension, allowing the measurement of the trace of a single laser shot [20–23]. Thus, all the pulses within a train can be measured, obtaining the full pulse train dynamics. However, this approach can be impractical under certain circumstances, specifically at low pulse intensity or high repetition rate. A low pulse intensity complicates the generation and detection of the nonlinear signal, while a high repetition rate would require a huge number of measurements and retrievals, requiring extensive computational time and power. Furthermore, increasing the repetition rate implies a shorter integration time to isolate a single pulse, requiring higher peak intensity, and even potentially exceeding the capabilities of typical spectrometers, whose minimum integration time is about 10 μs (maximum repetition rate of 100 kHz [24]).

Under these circumstances, the second approach, based on statistical analysis, proves to be more practical and efficient. Despite not capturing the full temporal evolution, the pulse train can be characterized by a base pulse and its variability within the train. This approach applies to scanning architecture, and to single-shot modality (in the case of multi-pulse integration). The retrieved information is enough to provide a representative profile of the pulse train and can be helpful to optimize it and to remove the instabilities, undesired in most cases [25]. For this purpose, FROG has previously been applied, modelling the train as a combination of a coherent artifact, the shortest repeatable structure within the train, and an average pulse, a longer pulse whose duration correlates with the degree of pulse variability [26–28]. Similarly, MIIPS was used to study instabilities in mode-locked semiconductor lasers [29]. Later, self-calibrating d-scan [9] was applied to quantify and correct instabilities within an Erbium-doped fiber laser. In this case, the discrepancy between the known dispersion of the compressor and the dispersion value retrieved by the algorithm provides the instability metric [30–32].

In this work, we analyze the impact of shot-to-shot pulse-shape fluctuations on the a-swing trace, providing a useful tool for the characterization of unstable pulse trains, since a-swing can be applied to diverse pulse parameters [11,33,34]. Section 2 presents an in-depth study of the a-swing trace, analyzing its Fourier components. This study is used here to identify instabilities and improve reconstructions under noisy regimes, as well as to understand how pulse information is encoded in the trace, potentially leading to refined retrieval strategies. Section 3 introduces a set of five pulse trains exhibiting varying degrees of instability. In Section 4, their traces are computed and then retrieved using a ptychographic algorithm [13]. Pulse-shape instabilities manifest as discrepancies between the simulated (or experimental) and retrieved frequency-dependent second and sixth Fourier coefficients of the trace (Eq. 5). Based on these discrepancies, we propose a metric to quantify the instabilities, whose performance is validated under noisy conditions. We show that a-swing is sensitive to pulse-shape fluctuations along the train even in the presence of random noise, establishing it as a suitable tool for the temporal characterization of both stable and unstable pulse trains.

## 2. Fourier analysis of a-swing traces

The a-swing setup scans the relative amplitude of two time-delayed replicas of an unknown pulse —assumed here to be horizontally polarized. This inline common-path generation is facilitated by a rotating MWP and a linear polarizer (LP). After this pulse manipulation, the second-harmonic generation (SHG) spectrum is produced in a nonlinear crystal (NL) and measured for each orientation of the MWP $\theta$, i.e., representing relative amplitude, yielding the 2D trace (Fig. 1). The spectral-domain representation of the pulse before the SHG is:

$$E_{\mathrm{LP}}(\omega,\theta) = E(\omega)\left[\cos^2\theta\, \mathrm{e}^{\mathrm{i}\,\rho_{\mathrm{f}}(\omega)} + \sin^2\theta\, \mathrm{e}^{\mathrm{i}\,\rho_{\mathrm{s}}(\omega)}\right] \quad (1)$$

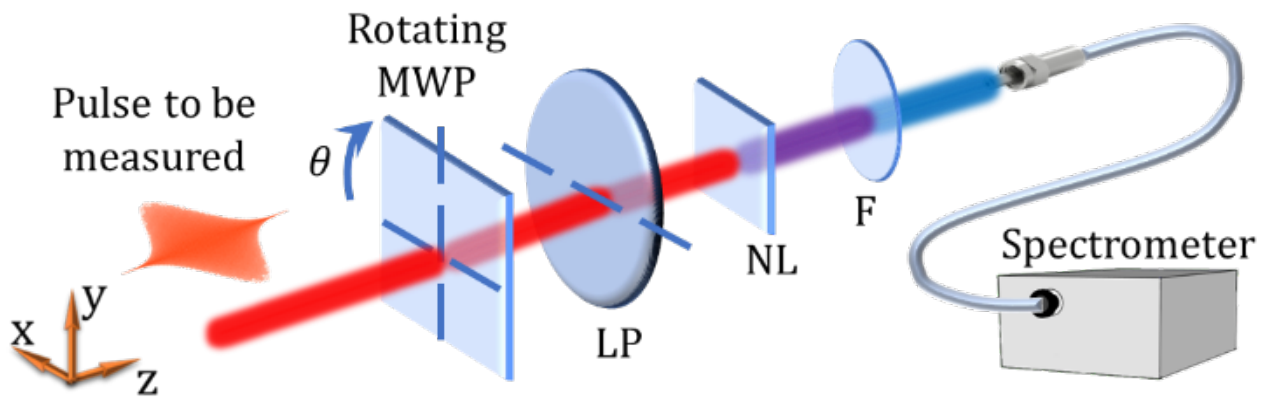

**Fig. 1.** Amplitude swing conventional experimental setup [4]. Pulse manipulation: multiple-order waveplate (MWP) and linear polarizer (LP). Nonlinear generation and detection: nonlinear crystal (NL), filter of the fundamental signal (F) and spectrometer.

where $E(\omega) = A(\omega)\,\mathrm{e}^{\mathrm{i}\varphi(\omega)}$ represents the input pulse, and $\rho_\mathrm{f}(\omega)$ and $\rho_\mathrm{s}(\omega)$ denote the phases due to the MWP fast and slow axes, respectively. The SHG of this pulse is subsequently calculated as the square of the temporal field profile:

$$E_\mathrm{SHG}(\omega,\theta) = \int \left( \int E_\mathrm{LP}(\omega',\theta)\,\mathrm{e}^{-\mathrm{i}\omega' t}\,\mathrm{d}\omega' \right)^2 \mathrm{e}^{+\mathrm{i}\omega t}\,\mathrm{d}t \tag{2}$$

which can be written as

$$E_\mathrm{SHG}(\omega,\theta) = E_\mathrm{ff}(\omega)\cos^4\theta + 2E_\mathrm{fs}(\omega)\cos^2\theta\sin^2\theta + E_\mathrm{ss}(\omega)\sin^4\theta \tag{3}$$

where $E_\mathrm{ff}(\omega)$ and $E_\mathrm{ss}(\omega)$ represent the SHG signals of the fast and slow replicas, respectively, and $E_\mathrm{fs}(\omega)$ the sum-frequency generation (SFG) between both. Then, the a-swing trace is defined as the modulus square of this SHG field:

$$\begin{aligned} S_\mathrm{SHG}(\omega,\theta) = |E_\mathrm{SHG}(\omega,\theta)|^2 = s_\mathrm{4f}(\omega)\cos^8\theta + s_\mathrm{3f,s}(\omega)\cos^6\theta\sin^2\theta \\ + s_\mathrm{2f,2s}(\omega)\cos^4\theta\sin^4\theta + s_\mathrm{f,3s}(\omega)\cos^2\theta\sin^6\theta + s_\mathrm{4s}(\omega)\sin^8\theta \end{aligned} \tag{4}$$

where the subscripts of the *s*-coefficients indicate the terms of the SHG pulse from which they originate. This expression can be expanded as a Fourier series:

$$S_\mathrm{SHG}(\omega,\theta) = a_0(\omega) + a_2(\omega)\cos 2\theta + a_4(\omega)\cos 4\theta + a_6(\omega)\cos 6\theta + a_8(\omega)\cos 8\theta \tag{5}$$

obtaining that the trace presents just even cosine terms up to order 8, since it is symmetric with respect to $\theta = 90^\mathrm{o}$ and it is periodic every $180^\mathrm{o}$. The equivalence between the *s*-coefficients and the Fourier coefficients is given by

$$\begin{pmatrix} s_\mathrm{4f}(\omega) \\ s_\mathrm{3f,s}(\omega) \\ s_\mathrm{2f,2s}(\omega) \\ s_\mathrm{f,3s}(\omega) \\ s_\mathrm{4s}(\omega) \end{pmatrix} = \begin{pmatrix} +1 & +1 & +1 & +1 & +1 \\ +4 & +2 & -4 & -14 & -28 \\ +6 & 0 & -10 & 0 & +70 \\ +4 & -2 & -4 & +14 & -28 \\ +1 & -1 & +1 & -1 & +1 \end{pmatrix} \begin{pmatrix} a_0(\omega) \\ a_2(\omega) \\ a_4(\omega) \\ a_6(\omega) \\ a_8(\omega) \end{pmatrix} \tag{6}$$

The frequency-dependent Fourier coefficients can be directly obtained from the trace, without knowing the pulse, thus we can also calculate the *s*-coefficients before the retrieval, which provides preliminary information about the pulse, such as the sign of the phase. Furthermore, we use the trace marginals $M_\omega(\omega)$ and $M_\theta(\theta)$, that result from integrating the trace across its two axes, and are related with the Fourier coefficients by:

$$\begin{aligned} M_\omega(\omega) &= \int_0^\pi S_\mathrm{SHG}(\omega,\theta)\,\mathrm{d}\theta = \sum_{n=0}^{4} \left( a_{2n}(\omega) \int_0^\pi \cos 2n\theta\ \mathrm{d}\theta \right) = \pi\, a_0(\omega) \\ M_\theta(\theta) &= \int_{-\infty}^{+\infty} S_\mathrm{SHG}(\omega,\theta)\,\mathrm{d}\omega = \sum_{n=0}^{4} \left( \cos 2n\theta \int_{-\infty}^{+\infty} a_{2n}(\omega)\,\mathrm{d}\omega \right) \end{aligned} \tag{7}$$

From these expressions, we conclude that the offset term $a_0(\omega)$ is proportional to the frequency marginal, as expected, since it is not modulated by $\theta$. We can also deduce that $a_2(\omega) = -a_6(\omega)$, since the intensity must be identical at the orientations where only a single replica exists, $\theta = 0, 90, 180$º (considering that the MWP relative dispersion is small within the pulse bandwidth). Furthermore, we can deduce the relation between the coefficients of two pulses with conjugated spectral phases, $a_2^{(\varphi)}(\omega) = -a_2^{(-\varphi)}(\omega)$, which means that the traces of these phase-conjugate pulses are shifted by 90º between them.

## 3. Simulation of the unstable pulse trains

To simulate the set of unstable pulse trains, a base pulse has been initially defined with a Gaussian spectrum centered at 800 nm, presenting a FTL duration of 50 fs, and a spectral phase given by $-3000$ fs$^2$ of group delay dispersion (GDD) and $+10^5$ fs$^3$ of third-order dispersion (TOD) (Fig. 2). These parameters were chosen to match realistic experimental conditions, while the observed qualitative behavior does not depend on their exact values. To this base pulse, we add random variations to the spectral phase, without modifying the spectrum, resulting in five trains of 101 different pulses each. A similar methodology to construct the pulse trains has been employed in previous works with FROG and d-scan [25,26,30,31]. Thus, each pulse within the train is defined by the spectrum of the base pulse and a spectral phase $\varphi_i(\omega)$ expressed as:

$$\varphi_i(\omega) = \frac{-3000\ \mathrm{fs}^2}{2}(\omega - \omega_0)^2 + \frac{10^5\ \mathrm{fs}^3}{6}(\omega - \omega_0)^3 + \alpha \cdot \mathcal{U}_\omega(-\pi, \pi) \tag{8}$$

where $\omega_0$ denotes the central frequency and $\alpha$ is the parameter governing the degree of instability. This instability is introduced via a function $\mathcal{U}_\omega$ that assigns independent random values to the phase of each frequency component. After this, the phase is smoothed by applying a super-Gaussian temporal filter (order 3, width 1200 fs) to remove noise at temporal delays far from the pulse peak. Following this procedure, we defined five pulse trains corresponding to $\alpha = [0.15, 0.30, 0.45, 0.60, 0.75]$ (Fig. 3). Tab. 1 summarizes the statistical parameters of each train in terms of the full width at half maximum (FWHM) and second-order central moment (SCM) pulse durations. The former reflects the duration of the most intense peak, while the latter, defined as twice the square root of the SCM of the intensity $\Delta t_{2\sigma} = 2\sigma_t$, with $\sigma_t^2 = \langle t^2 \rangle - \langle t \rangle^2$, considers the secondary temporal structure of the pulse, i.e., it serves as a measure of the temporal spreading (it corresponds to the 1/e$^2$ width in the case of a Gaussian profile). The train preserves the main features of the base pulse for $\alpha \leq 0.45$, while for greater values, especially for $\alpha = 0.75$, this structure presents significant degradation.

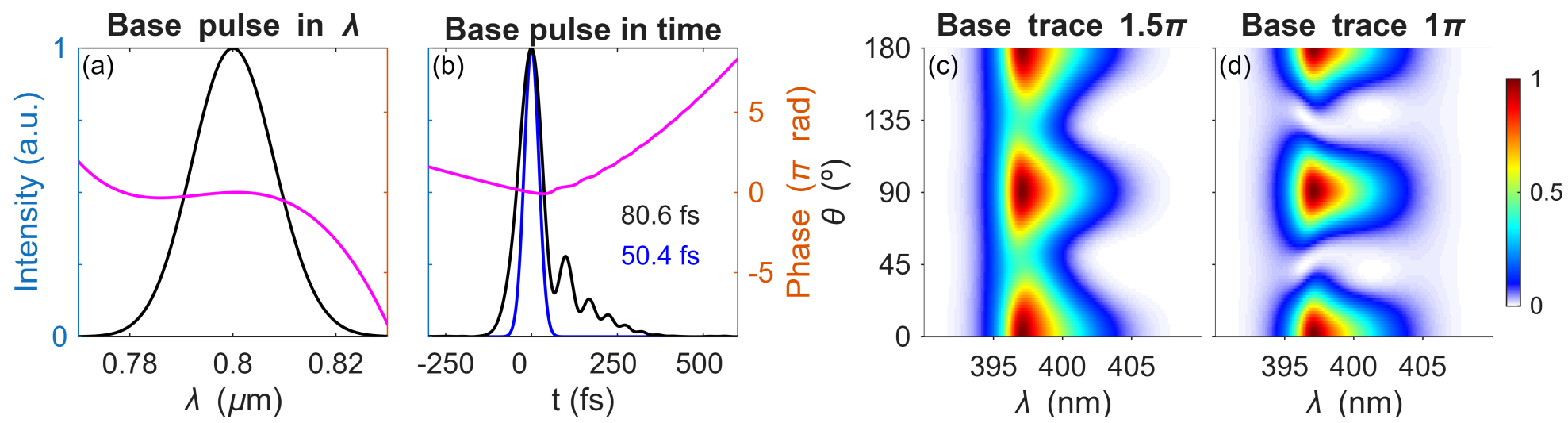

**Fig. 2.** Base pulse in: (a) wavelengths; (b) time, with the FTL pulse in blue and the FWHM durations indicated. Black and blue: intensity; magenta: phase. Simulated traces for MWP phase retardation equal to: (c) 1.5$\pi$; (d) $\pi$. The MWP delay is 50 fs in both cases.

**Table 1.** Statistics of pulse durations of the different pulse trains. The FWHM and SCM durations of the base pulse ($\boldsymbol{\alpha = 0}$) are 80.6 and 188.7 fs, respectively.

| | $\alpha = 0.15$ | | $\alpha = 0.30$ | | $\alpha = 0.45$ | | $\alpha = 0.60$ | | $\alpha = 0.75$ | |
|---|---|---|---|---|---|---|---|---|---|---|
| | **FWHM** | **SCM** | **FWHM** | **SCM** | **FWHM** | **SCM** | **FWHM** | **SCM** | **FWHM** | **SCM** |
| **Mean (fs)** | 81.5 | 191.9 | 81.6 | 213.3 | 81.0 | 263.1 | 82.1 | 367.58 | 84.22 | 516.72 |
| **Mean / base** | 1.01 | 1.02 | 1.01 | 1.13 | 1.01 | 1.39 | 1.02 | 1.95 | 1.05 | 2.74 |
| **std / mean (%)** | 3.62 | 2.02 | 7.49 | 5.29 | 10.90 | 10.73 | 18.41 | 19.26 | 34.57 | 19.2 |
| **Norm var (%)** | 0.13 | 0.04 | 0.56 | 0.28 | 1.19 | 1.15 | 3.39 | 3.71 | 11.95 | 3.69 |

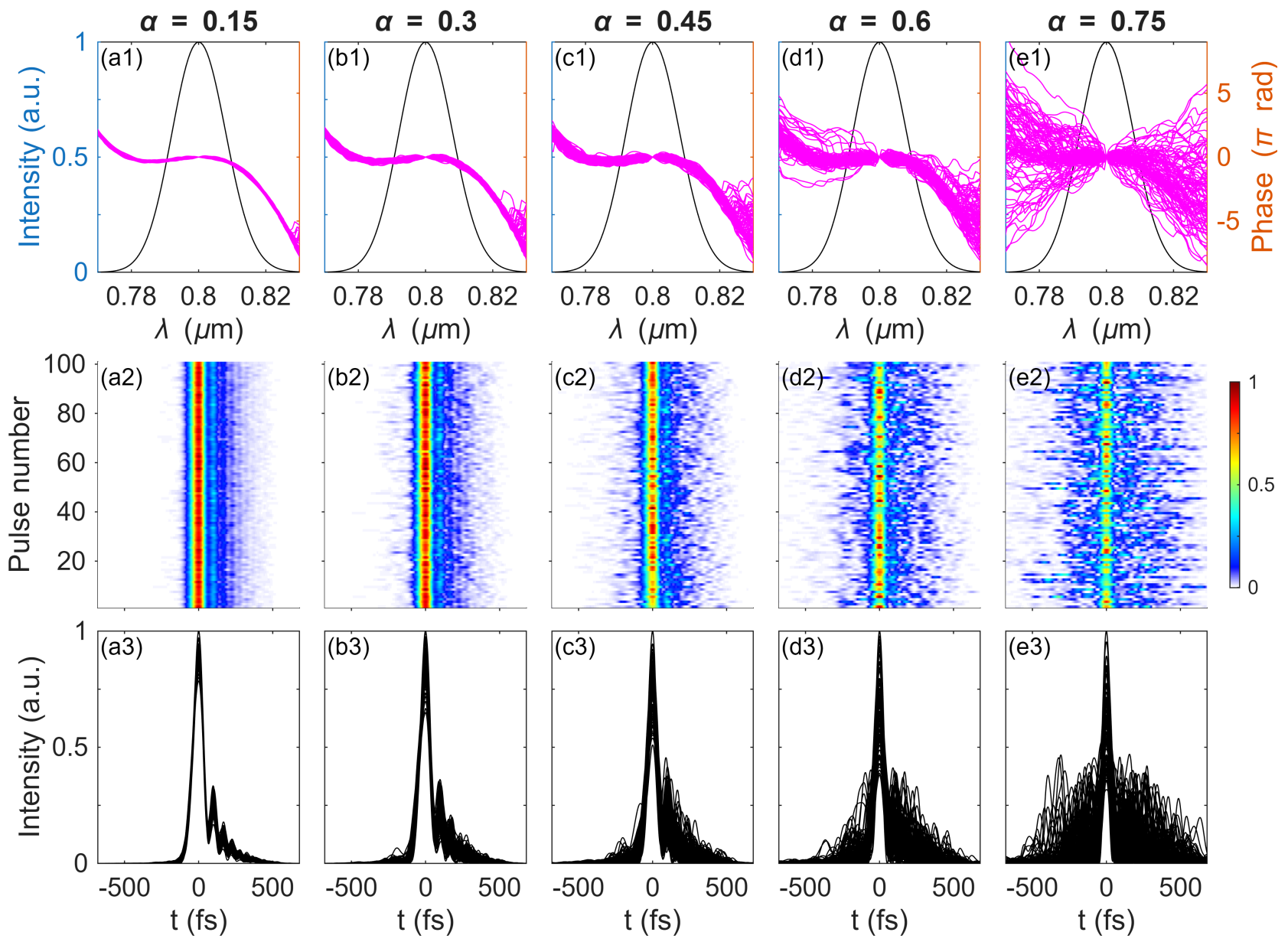

**Fig. 3.** Unstable pulse trains for: (a) $\alpha = 0.15$; (b) $\alpha = 0.3$; (c) $\alpha = 0.45$; (d) $\alpha = 0.6$; (e) $\alpha = 0.75$. Row 1: spectrum (black) and spectral phases (magenta); Rows 2 and 3: temporal intensities (indicated by the color bar in row 2; black curves in row 3).

## 4. Results and discussion

In this section, we present the simulated a-swing traces of the previously mentioned pulse trains (Fig. 3), using two 1.6-mm-thick MWPs of α-quartz. The birefringence and thickness of the plate produce a 50-fs delay, corresponding to the FTL pulse duration, which is the optimal value for a-swing retrievals of stable pulse trains [4,33]. In addition, the MWP introduces a multi-order retardation. To indicate its value, the $2\pi m$ offset is removed, and only the zero-order relative phase is reported in the phase retardation, which, unless specified otherwise, refers to the central frequency (the dephasing of the MWP is chromatic).

The results are presented for MWP phase retardations of π and 1.5π, corresponding to half-wave plate (HWP) and quarter-wave plate (QWP) operation, respectively, although the observed behavior and the proposed metric hold for arbitrary phase retardations. The simulated traces have been reconstructed using the ptychographic algorithm developed for a-swing in Ref. [13]. We present the retrieved pulses for different degrees of phase instability, and we propose a metric to quantify these instabilities, allowing to distinguish between stable and unstable pulse trains. Furthermore, we analyze the robustness of this metric against random noise and propose a method to distinguish the contributions of instabilities and noise.

### *4.1 Amplitude swing retrieval of traces from unstable pulse trains*

Here, we show the a-swing simulated and retrieved traces for phase retardations π and 1.5π, for different degrees of instability, controlled by $\alpha$ (Eq. 8). As $\alpha$ increases, the deviation between the base trace and the trace of the unstable pulse train grows, as expected. For $\alpha = 0.15$ and $\alpha = 0.3$, the trace barely changes, thus the algorithm retrieves the base pulse (Fig. 4 and 5, row 1). For higher instability values, the retrieved pulses progressively lose their secondary structure, while retrieving well the main pulse peak (Fig. 4 and 5, row 4).

Even in the case of maximum instability, where the simulated traces significantly differ from the base trace, the $G$ error (Eq. 9) remains below 2%, as the main differences arise in the zones of null intensity (Fig. 4 and 5, row 3). Thus, this error is not an adequate metric to detect and quantify instabilities since one may confuse a stable train of clean pulses with an unstable pulse train of very structured pulses. Fortunately, instabilities clearly manifest as discrepancies between the simulated and retrieved Fourier coefficients $a_2(\omega)$ and $a_6(\omega)$, as shown in the next subsection.

$$G = \frac{1}{\sqrt{N_\omega N_\theta}} \sqrt{\sum_{\omega,\theta} \left(S_{\mathrm{SHG}}^{\mathrm{retr}}(\omega,\theta) - S_{\mathrm{SHG}}^{\mathrm{sim}}(\omega,\theta)\right)^2} \tag{9}$$

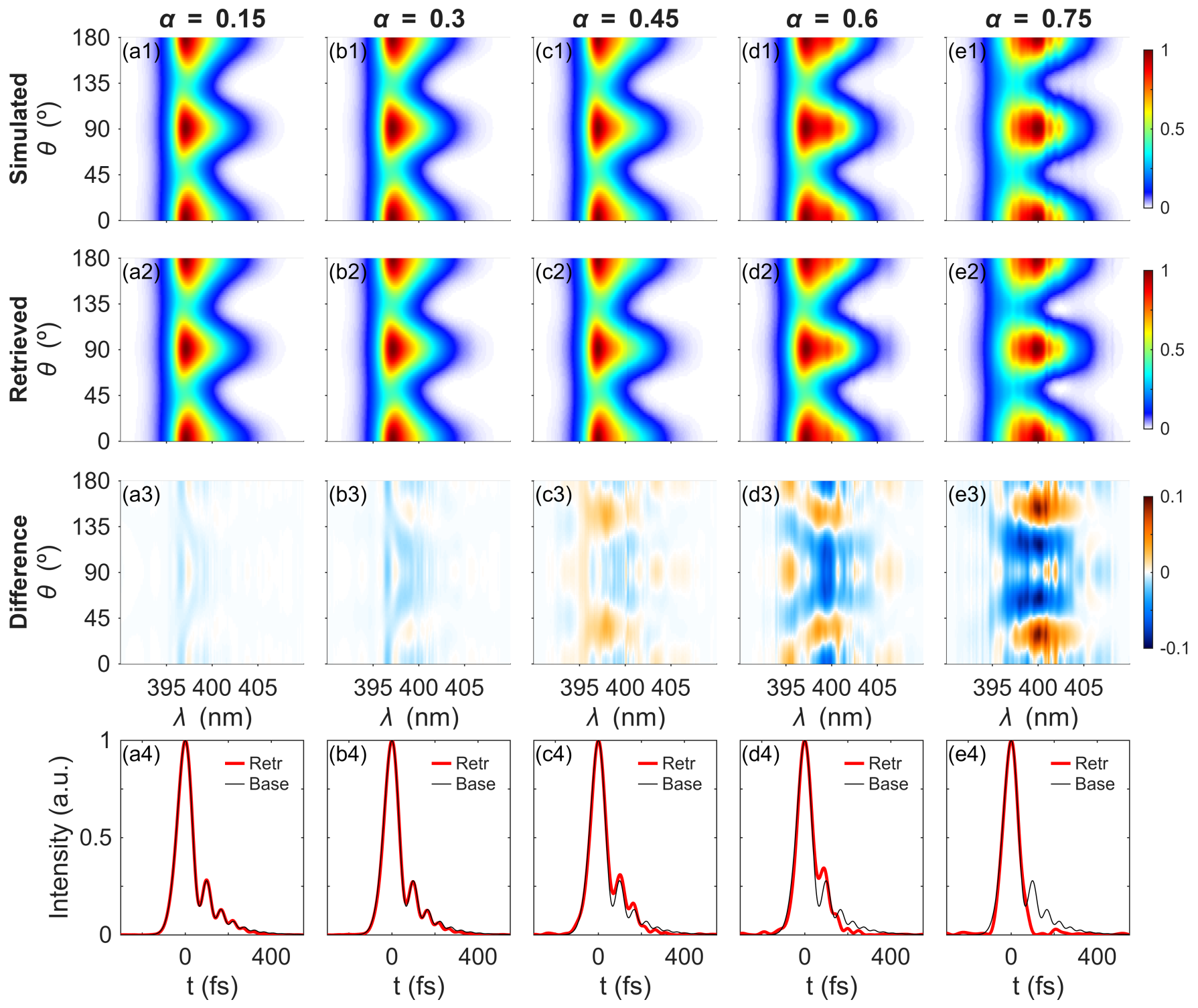

**Fig. 4.** Retrievals for MWP operating as a QWP, with phase retardation of 1.5π: (a) $\alpha = 0.15$; (b) $\alpha = 0.3$; (c) $\alpha = 0.45$; (d) $\alpha = 0.6$; (e) $\alpha = 0.75$. Row 1: simulated traces; Row 2: retrieved traces; Row 3: difference; Row 4: retrieved pulses in time domain.

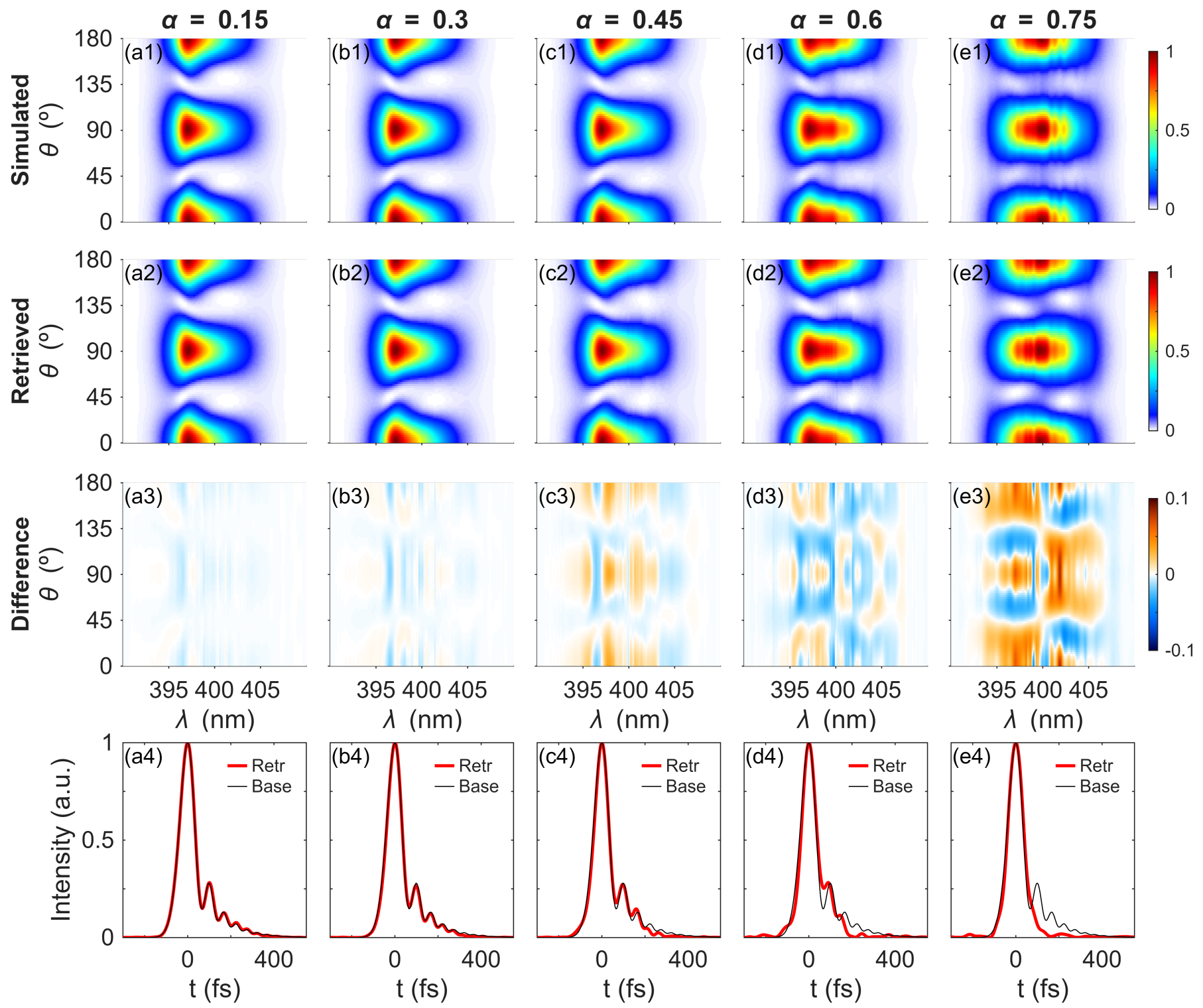

**Fig. 5.** Retrievals for MWP operating as a HWP, with phase retardation of π: (a) $\alpha = 0.15$; (b) $\alpha = 0.3$; (c) $\alpha = 0.45$; (d) $\alpha = 0.6$; (e) $\alpha = 0.75$. Row 1: simulated traces; Row 2: retrieved traces; Row 3: difference; Row 4: retrieved pulses in the temporal domain.

The minimum-intensity regions in the retrieved traces are more pronounced than those in the simulated traces from the unstable pulse train because the retrieval assumes a single, stable pulse. Since the traces of the different pulses within the unstable train present null intensity at different positions, their incoherent superposition reduces the zero-intensity areas of the trace. This behavior could help identify the presence of instabilities qualitatively, especially for a phase retardation of π, for which there are more minimum-intensity regions.

For an MWP phase retardation of 1.5π (or, equivalently, 0.5π), the variation of the second Fourier coefficient leads to a shift of the minima of the relative amplitude marginal. In contrast, in the case of symmetric spectral amplitude and phase, an MWP phase retardation of π results in a marginal that is periodic every 90º (equivalently, exhibiting symmetry with respect to 45º and 135º); as a result, the integral of the second and sixth Fourier coefficients is null. In our case, the TOD slightly breaks that symmetry and thus the integral of said Fourier coefficients are approximately zero (Fig. 6, Row 2).

In addition to the noise-free traces, the retrievals are performed after adding uniform random noise of 4% and 8% to the simulated traces of the unstable pulse trains. These retrievals are improved using the Fourier decomposition of the trace by removing the noise along the $\theta$ axis (see Supplement 1, Fig. S1). In the next subsection, we propose a metric to quantify the instabilities and assess its performance both for noiseless and noisy traces.

### 4.2 Quantification of pulse-train instabilities and noise robustness

As said before, the difference between the retrieved and simulated second Fourier coefficient grows when increasing the degree of instability (Fig. 6), serving to identify pulse-train instabilities. We propose the metric Γ to quantify them:

$$\Gamma = \frac{\int \left| a_2^{\mathrm{retr}}(\omega) - a_2^{\mathrm{sim}}(\omega) \right| \, \mathrm{d}\omega}{\int \left| a_2^{\mathrm{retr}}(\omega) \right| \, \mathrm{d}\omega} \tag{10}$$

which is proportional to the area between the simulated (or experimental) and retrieved second Fourier coefficients. This parameter yields a value of 0 in the absence of instabilities, and it approaches 1 as the instabilities increase.

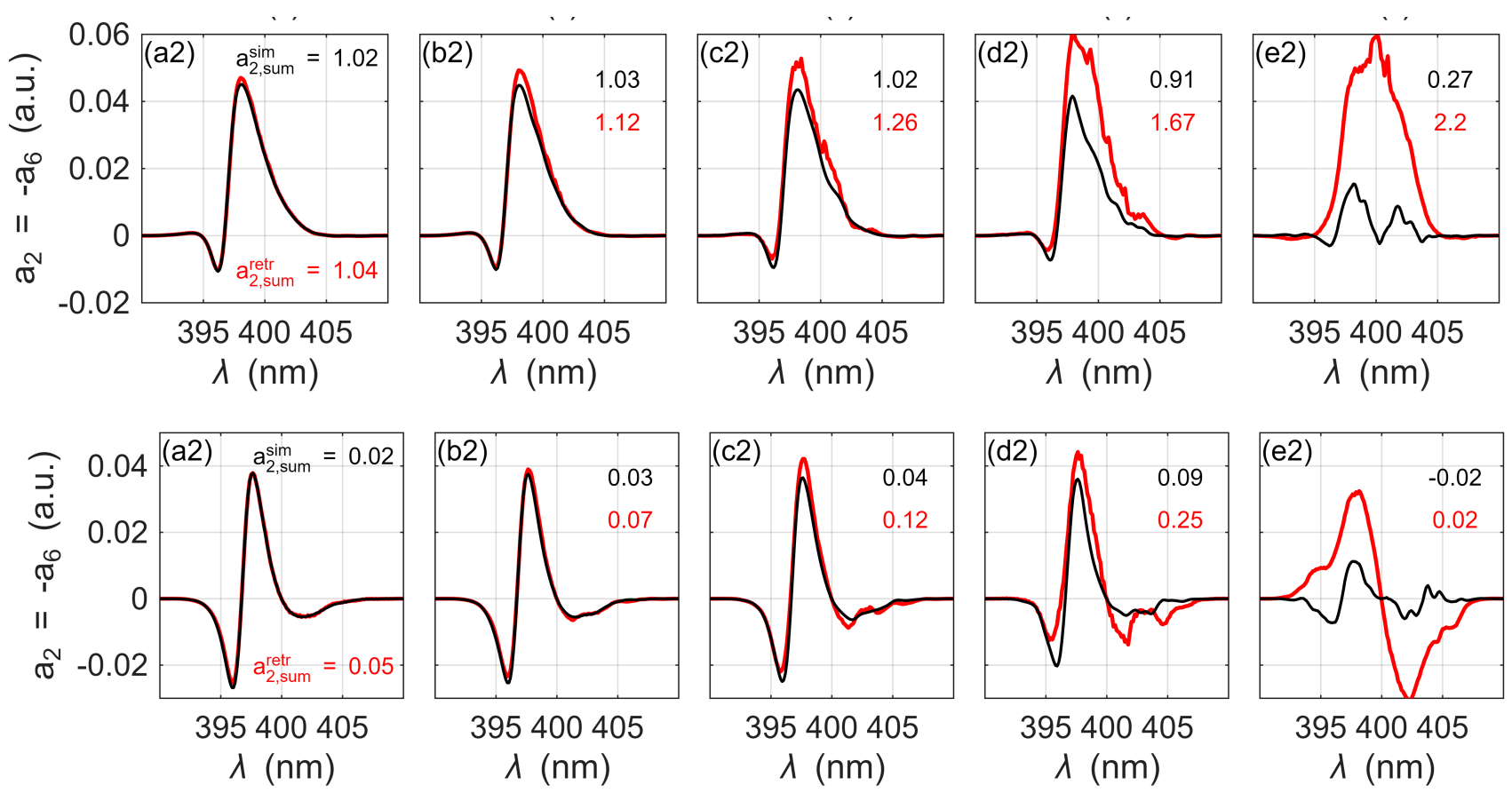

**Fig. 6.** Effect of pulse-train instabilities for a phase retardation of 1.5π (first row) and π (second row) in the simulated (black) and retrieved (red) second Fourier coefficient $a_2(\omega)$ for: (a) $\alpha = 0.15$; (b) $\alpha = 0.3$; (c) $\alpha = 0.45$; (d) $\alpha = 0.6$; (e) $\alpha = 0.75$.

The proposed metric Γ increases linearly with the average SCM pulse duration $\langle \Delta t_{2\sigma} \rangle$ for both MWP phase retardations (Fig. 7a,b), as desired to quantify the pulse-train instabilities. To evaluate the robustness of this metric, we analyze its behavior under the presence of random noise, using noise levels well above the typical experimental conditions. The full set of simulated and retrieved traces are shown in Supplement 1, jointly with their Fourier coefficients.

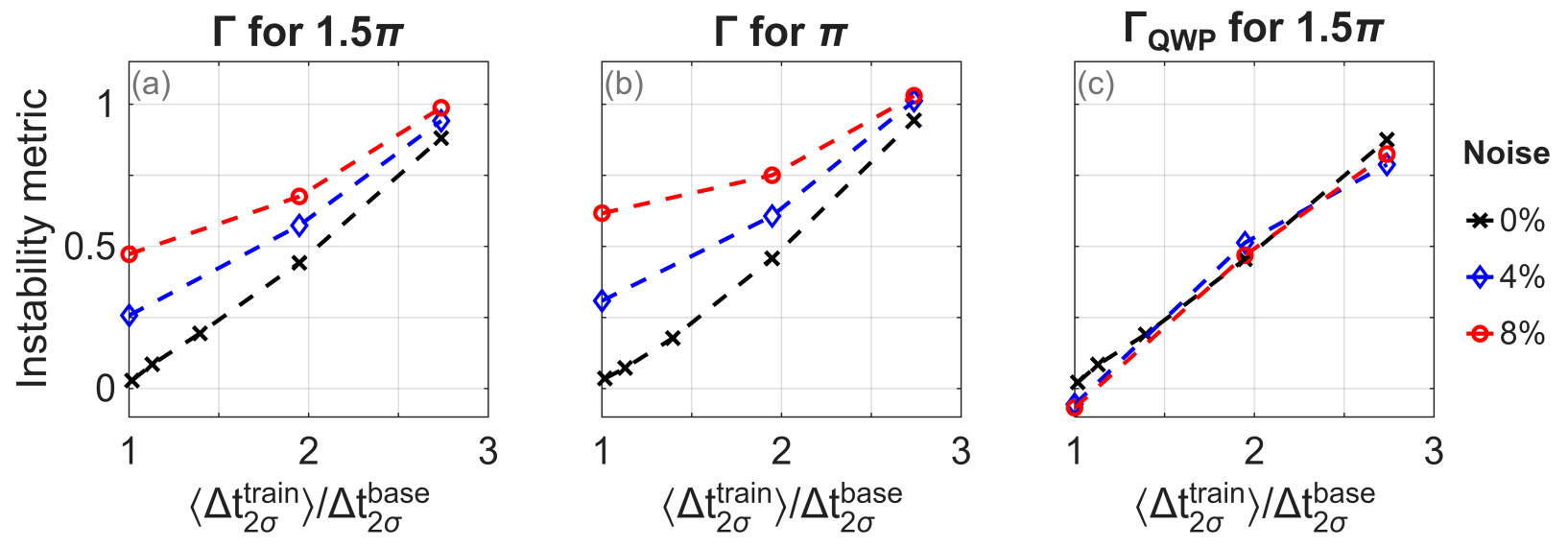

**Fig. 7.** Instability metrics as a function of the average of the SCM pulse duration and noise level: (a) Γ for a phase retardation of 1.5π; (b) for a phase retardation of π; (c) $\Gamma_{\mathrm{QWP}}$ for a phase retardation of 1.5π. Black: 0% noise; Blue: 4%; Red: 8%.

Ideally, the metric should not yield false indications of instability in the presence of noise alone, while remaining sensitive to the presence of instabilities. The metric Γ maintains its dependence with the degree of instability, but its offset increases and its slope decreases with the noise level. For small degrees of instability, the random noise increases the area between the simulated and retrieved Fourier coefficients, in contrast with high instability, when it barely affects that area. Thus, $\Gamma$ presents contribution from pulse-train instabilities and from random noise, $\Gamma = \Gamma^{\text{inst}} + \Gamma^{\text{noise}}$, and it is required to separate them, even though their effects can be clearly distinguished qualitatively.

To do this, for an MWP phase retardation of $1.5\pi$, we can use the metric without the absolute values:

$$\Gamma_{\text{QWP}} = 1 - \frac{\int a_2^{\text{sim}}(\omega)\,\mathrm{d}\omega}{\int a_2^{\text{retr}}(\omega)\,\mathrm{d}\omega} \tag{10}$$

which remains robust against noise, exhibiting the same dependence with the degree of instability for different noise levels (Fig. 9a). This metric is not valid for MWP phase retardations close to $\pi$ because the integral of the second Fourier coefficient is close to zero.

For arbitrary phase retardations, one possibility is to compute the integral $\int \left( \left| a_2^{\text{retr}}(\omega) \right| - \left| a_2^{\text{sim}}(\omega) \right| \right) \mathrm{d}\omega$ as a complimentary metric. This difference considers the sign of the area, resulting in a close-to-zero value when the difference is caused by random noise. Another approach could be to smooth the Fourier coefficients by removing the high-frequency noise before the comparison, but we prefer to use the raw data to avoid introducing artifacts.

A more elegant solution is the autocorrelation of the difference $a_2^{\text{retr}}(\omega) - a_2^{\text{sim}}(\omega)$. It is well known that the autocorrelation of white noise corresponds to a Dirac delta function. Consequently, pure instabilities produce broad smooth autocorrelations, while pure noise produce autocorrelations with a narrow sharp peak. When both are present, the autocorrelation exhibits both features: a narrow sharp superimposed on a broad smooth background (Fig. 8). These two features can be separated, allowing quantification of the individual contributions of random noise $\Gamma^{\text{noise}}$and pulse-train instabilities $\Gamma^{\text{inst}}$, i.e., which fraction of $\Gamma$ corresponds to each contribution. We calculate the ratio $\Gamma^{\text{noise}}/\Gamma$ as the ratio between the height of the narrow sharp peak to the total height of the autocorrelation, obtaining the contribution of instabilities as $\Gamma^{\text{inst}} = \Gamma(1 - \Gamma^{\text{noise}}/\Gamma)$, which maintains its dependence with the degree of instability in the presence of high noise levels (Fig. 9b,c).

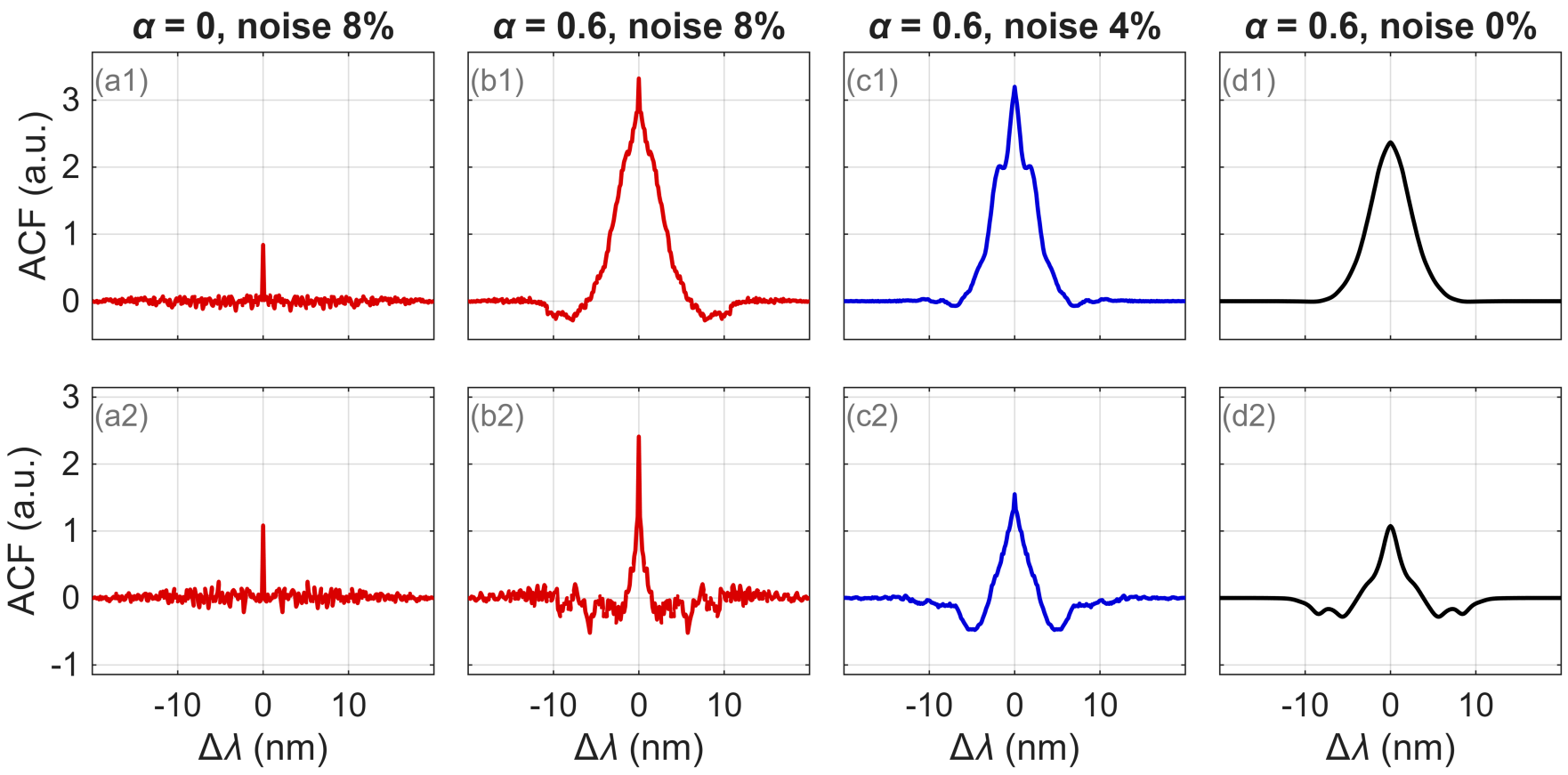

**Fig. 8.** Autocorrelation functions (ACF) of the difference $a_2^{\text{retr}}(\omega) - a_2^{\text{sim}}(\omega)$ for a phase retardation of $1.5\pi$ (row 1) and $\pi$ (row 2), for different levels of noise and degrees of instability: (a) stable pulse train with 8% noise, exhibiting only a narrow sharp peak, indicating pure noise; (b) noise of 8% and $\alpha = 0.6$; (c) noise of 4% and $\alpha = 0.6$; (d) noiseless trace and $\alpha = 0.6$, no sharp peak is on top of the broad function, indicating pure instabilities.

As discussed above, noise reduces the slope of the instability metric with respect to the degree of instability, thereby reducing its sensitivity to instabilities when increasing the noise level. This defines a detection threshold of degree of instability for a given noise level. Even for a noise level of 8%, well above than the typical experimental noise level, instabilities corresponding to $\alpha = 0.6$ remain detectable for both MWP phase retardations, with 1.5π providing higher sensitivity than π.

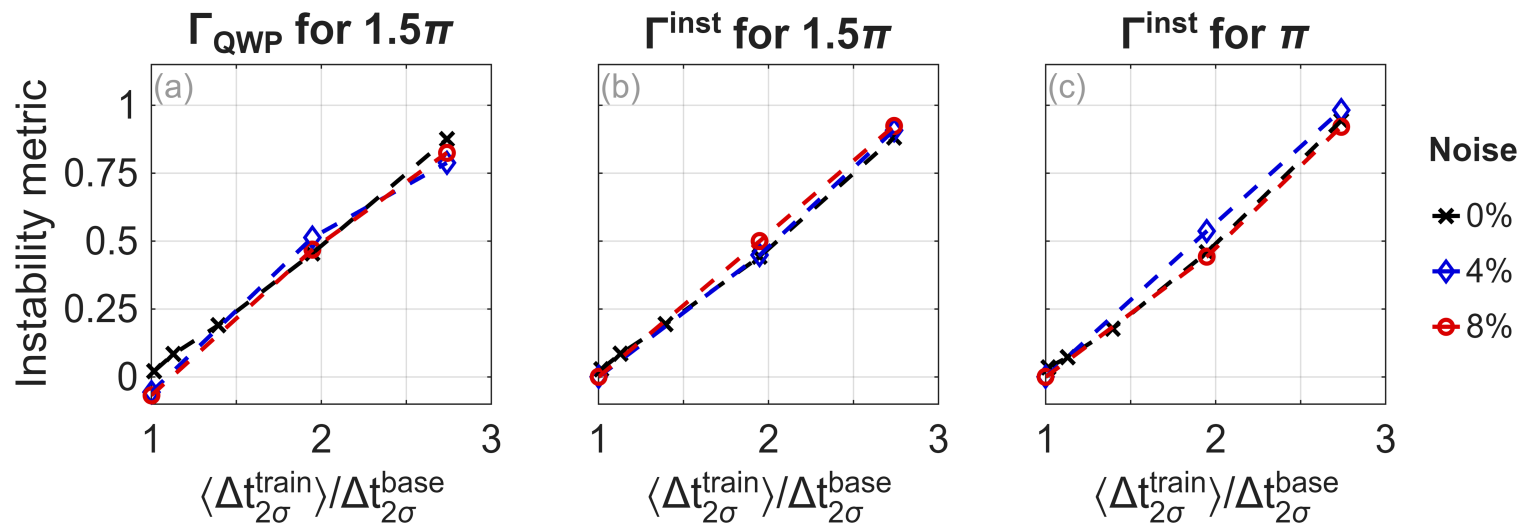

**Fig. 9.** Instability metrics after removing the noise contribution as a function of the average of the SCM pulse duration: (a) Γ for a phase retardation of 1.5π; (b) for a phase retardation of π; (c) $\Gamma_{QWP}$ for a phase retardation of 1.5π. Black: 0% noise; Blue: 4%; Red: 8%.

## 5. Conclusion

We have shown how pulse-train instabilities manifest in the amplitude swing trace, proposing a strategy to identify them and a metric to quantify them, as well as a procedure to qualitatively and quantitatively distinguish the effect of instabilities from random noise in the trace.

First, we have explored the constituent terms that compose the trace, decomposing it as a Fourier series with only even cosine terms up to eighth order. Then, we simulated a set of five pulse trains characterized by varying degrees of spectral phase instabilities. The traces of these pulse trains were simulated and retrieved, revealing that the second- and sixth-order frequency-dependent terms, $a_2(\omega)$ and $a_6(\omega)$, which are opposite, are modified by the presence of instabilities. Therefore, pulse-train instabilities can be identified and quantified by analyzing the discrepancies between these simulated (or experimental) and retrieved Fourier coefficients. Based on this, we propose the instability metric Γ, based on the area between the simulated and retrieved Fourier coefficients. Finally, we evaluate the metric performance under the presence of noise, finding that Γ presents contributions both from noise and instabilities. To separate both effects, we propose the autocorrelation of the difference of the Fourier coefficients to quantify their individual contributions, thereby quantifying pulse-train instabilities even under the presence of high noise levels. As an alternative, we propose the metric $\Gamma_{QWP}$, which is applicable when the phase retardation of the rotating multi-order waveplate is quarter-wave and it remains invariant under the presence of noise.

This analysis is especially useful for low-intensity, high-repetition-rate systems, where measuring every individual pulse is highly impractical. While this study focuses on scalar pulses with constant linear polarization using the conventional amplitude swing configuration, the transition to the generalized setup, or the inclusion of vector pulses with time-dependent polarization, reveal a more complex trace structure. In such cases, additional terms emerge within the signal, necessitating further investigation into future work.

In conclusion, we demonstrate that amplitude swing can distinguish between stable and unstable pulse trains, providing a simple and versatile tool to characterize ultrashort pulse trains presenting temporal profiles that vary shot to shot.

**Funding.** Ministerio de Ciencia, Innovación y Universidades (PID2023-149836NB-I00); Consejería de Educación, Junta de Castilla y León (SA108P24, Escalera de Excelencia CLU-2023-1-02); European Regional Development Fund.

**Acknowledgments.** C. Barbero acknowledges the support from Junta de Castilla y León and Fondo Social Europeo Plus through their Ph. D. grant program.

**Disclosures.** B. Alonso and Í. J. Sola have patent # WO/2021/123481 pending to University of Salamanca.

**Data availability.** Data underlying the results presented in this paper are not publicly available at this time but may be obtained from the authors upon reasonable request.

## QUANTIFICATION OF PHASE INSTABILITIES IN AMPLITUDE SWING FOR ULTRASHORT PULSE TRAIN CHARACTERIZATION: SUPPLEMENTAL DOCUMENT

Here, we show the simulated amplitude swing (a-swing) [1,2] traces of multiple sets of pulse trains, for different degrees of shot-to-shot pulse-shape instabilities and levels of noise. The noisy traces are computed by adding a random value to each point of the trace:

$$S_{\mathrm{SHG}}^{\mathrm{noisy}}(\omega_i,\theta_j) = S_{\mathrm{SHG}}(\omega_i,\theta_j) + \max\big(S_{\mathrm{SHG}}(\omega,\theta)\big)\cdot\mathcal{U}(-\eta,+\eta) \qquad \text{(S1)}$$

where $\mathcal{U}$ is a function that generates a random value between $-\eta$ and $+\eta$, being $\eta$ the noise level. The rms noise level corresponds to $\eta/\sqrt{3}$.

We simulate the a-swing traces for phase retardations of π and 1.5π, for instability values of 0, 0.6 and 0.75, and for noise levels of 0, 0.04 and 0.08. Before pulse retrieval, the noise along the $\theta$ axis is removed. To do this, we use the Fourier analysis of the a-swing signal. The traces are decomposed in their Fourier terms, arising more terms than those expected from the theoretical modulation, so a synthetic trace is built using Eq. 5, removing the terms that appear due to noise. For pulse retrieval, we used the ptychographic a-swing algorithm [3].

Fig. S1 shows the retrieved pulses from noisy and synthetic traces of a stable pulse train, simulated for a phase retardation of π. When the retrieval is performed using the synthetic traces, the agreement between the retrieved and simulated pulses is better than when is performed using the raw noisy traces. Therefore, the Fourier study of the a-swing trace further enhances the robustness of a-swing against noise and facilitates the subsequent quantification of pulse-train instabilities.

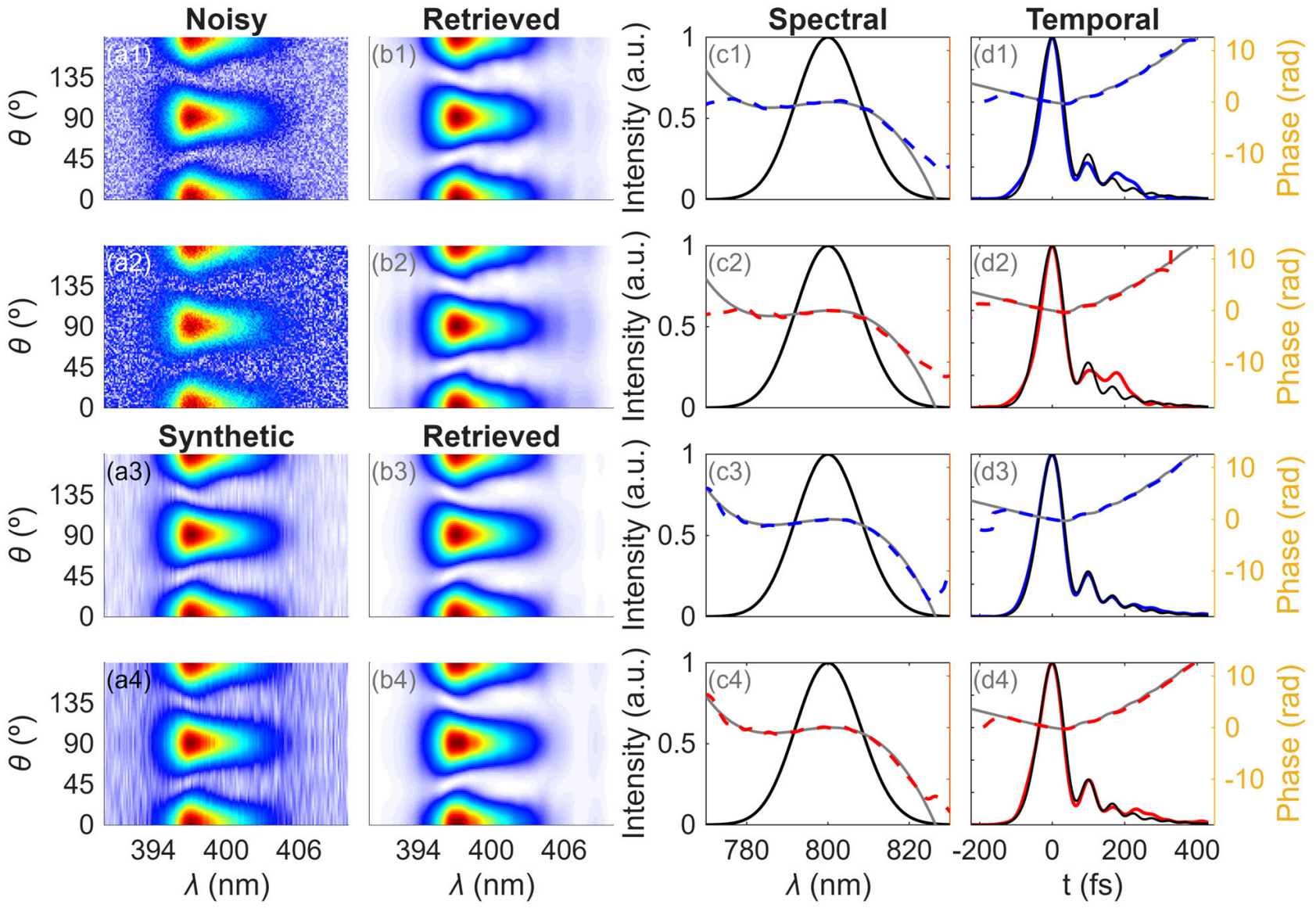

Fig. S1. A-swing retrieval of stable pulse trains under noisy conditions, directly from the noisy trace (rows 1 and 2), and after removing noise in the θ axis. (a) Noisy or synthetic a-swing traces; (b) retrieved traces; (c) spectrum (black), and simulated (gray) and retrieved (dashed red and blue) spectral phase; (d) simulated temporal intensity (black) and phase (gray), and retrieved temporal intensity (solid red and blue) and phase (dashed red and blue). Blue: 4% noise; red: 8% noise.

Fig. S2 shows the autocorrelation of the function $a_2^{\mathrm{retr}}(\omega) - a_2^{\mathrm{sim}}(\omega)$, computed from the noiseless traces for different degrees of pulse-train instability (row 2 of Fig. 5 and Fig. 7). The noisy, synthetic after noise in $\theta$ removed, and retrieved a-swing traces are shown in Fig. S3 for a phase retardation of 1.5π, and in Fig. S4 for a phase retardation of π. In addition, these figures display the corresponding simulated and retrieved second Fourier coefficients, and the autocorrelation of their difference.

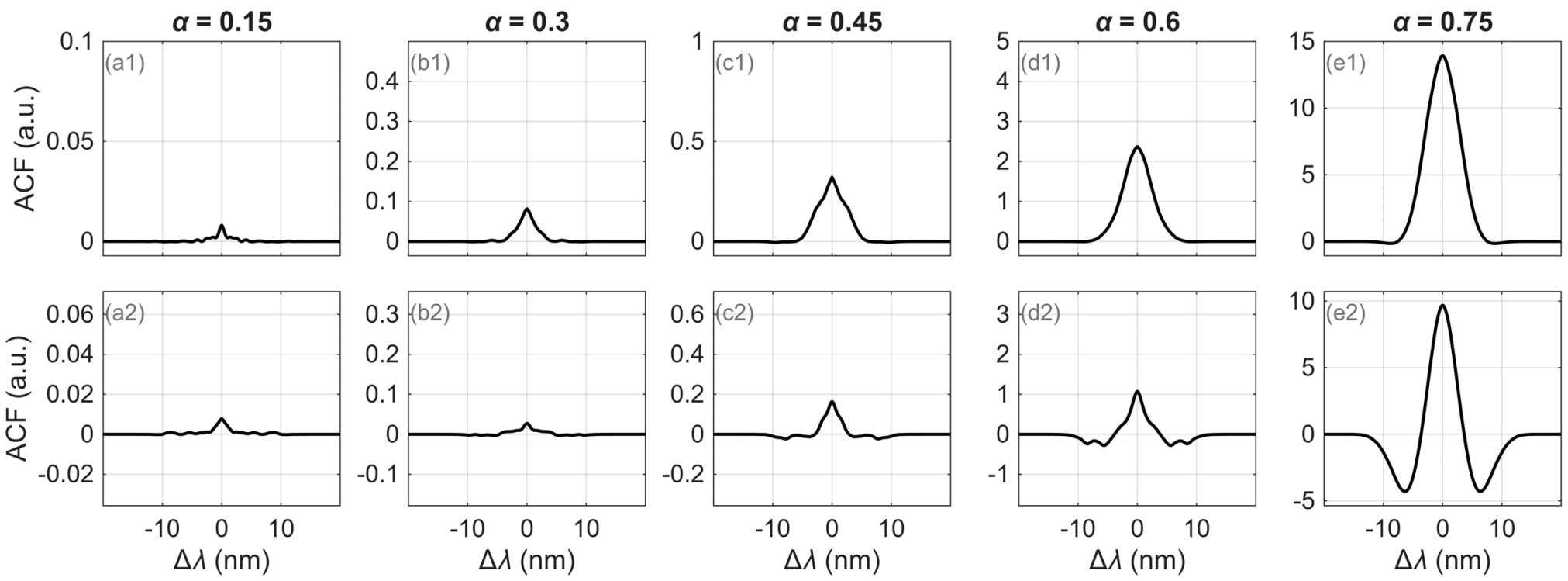

Fig. S2. Autocorrelation functions (ACF) of the difference $a_2^{\mathrm{retr}}(\omega) - a_2^{\mathrm{sim}}(\omega)$, for a phase retardation of 1.5π (row 1) and π (row 2), for different degrees of instability: (a) $\alpha = 0.15$; (b) $\alpha = 0.3$; (c) $\alpha = 0.45$; (d) $\alpha = 0.6$; (e) $\alpha = 0.75$.

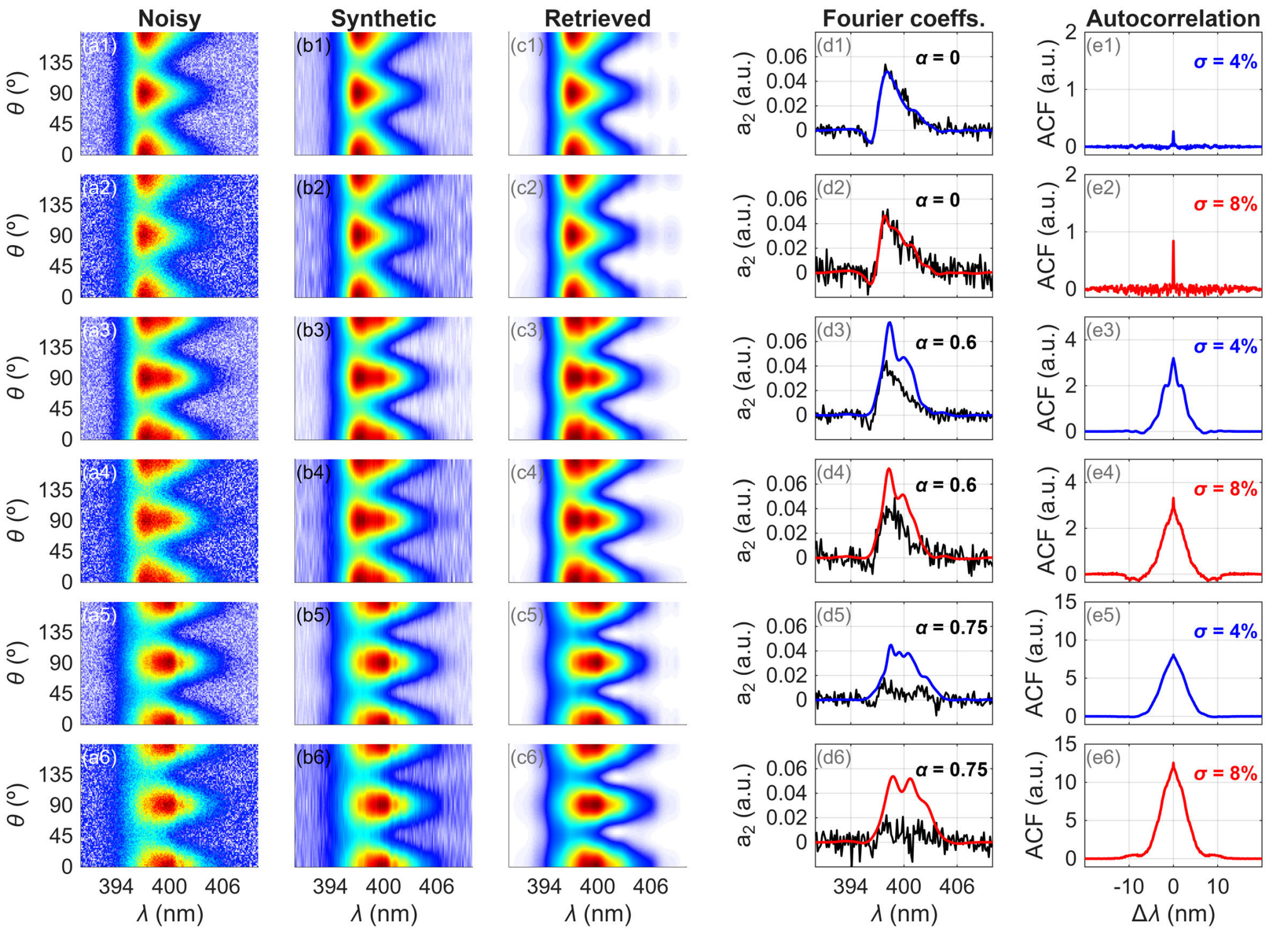

Fig. S3. A-swing retrievals for a phase retardation of 1.5π, for different degrees of instability and noise levels: (a) simulated noisy traces; (b) synthetic traces after removing the noise along the $\theta$ axis; (c) retrieved traces; (d) simulated (black) and retrieved (red and blue) second Fourier coefficients; (e) autocorrelation of the difference between the curves of panel (d).

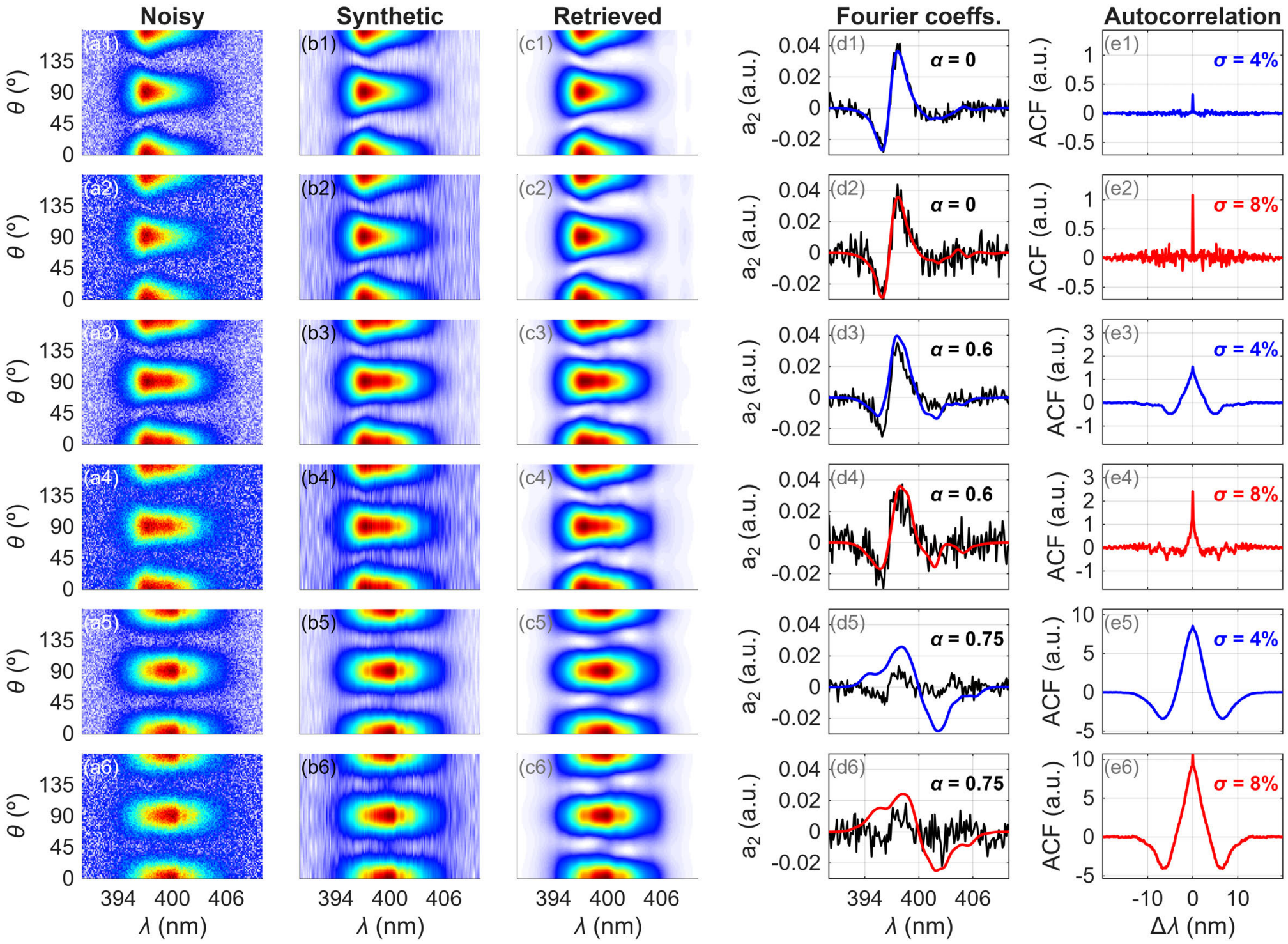

Fig. S4. A-swing retrievals using a phase retardation of π, for different degrees of instability and noise levels: (a) simulated noisy traces; (b) synthetic traces after removing the noise along the $\theta$ axis; (c) retrieved traces; (d) simulated (black) and retrieved (red and blue) second Fourier coefficients; (e) autocorrelation of the difference between the curves of panel (d).